\documentclass[conference]{IEEEtran}
\usepackage{amsmath}
\usepackage{url}
\usepackage{hyperref}
\usepackage{graphicx}
\usepackage[utf8]{inputenc}
\usepackage{tabularx}

\makeatletter

\def\ps@IEEEtitlepagestyle{
  \def\@oddfoot{\mycopyrightnotice}
  \def\@evenfoot{}
}
\def\mycopyrightnotice{
  {\footnotesize 979-8-3315-8310-1/25/\$31.00~\copyright~2025 IEEE\hfill} 
  \gdef\mycopyrightnotice{}
}

\@ifundefined{showcaptionsetup}{}{
 \PassOptionsToPackage{caption=false}{subfig}}
\usepackage{subfig}
\makeatother

\usepackage{eso-pic}
\newcommand\AtPageUpperMyright[1]{\AtPageUpperLeft{
 \put(\LenToUnit{0.5\paperwidth},\LenToUnit{-1cm}){
     \parbox{0.5\textwidth}{\raggedleft\fontsize{9}{11}\selectfont #1}}
 }}
\newcommand{\conf}[1]{
\AddToShipoutPictureBG*{
\AtPageUpperMyright{#1}
}
}

\begin{document}

\title{Comparison of Multiple Classifiers for Android Malware Detection with Emphasis on Feature Insights Using CICMalDroid 2020 Dataset}
\conf{2025 IEEE 7th International Conference on Sustainable Technologies for Industry 5.0 (STI 2025), 11-12 December, Dhaka} 

\author{\IEEEauthorblockN{Md Min-Ha-Zul Abedin\IEEEauthorrefmark{1}, Tazqia Mehrub\IEEEauthorrefmark{2}}
\IEEEauthorblockA{\IEEEauthorrefmark{1}Department of Biosystems Engineering, Auburn University, Auburn AL 36849, USA}
\IEEEauthorblockA{\IEEEauthorrefmark{2}Independent Researcher, Dhaka, Bangladesh}
\IEEEauthorblockA{Email: mza0288@auburn.com, etazkia@gmail.com}
}

\maketitle
\begin{abstract}
Accurate Android malware detection was critical for protecting users at scale. Signature scanners lagged behind fast release cycles on public app stores. We aimed to build a trustworthy detector by pairing a comprehensive dataset with a rigorous, transparent evaluation, and to identify interpretable drivers of decisions. We used CICMalDroid2020, which contained 17,341 apps across Benign, Adware, Banking, SMS malware, and Riskware. We extracted 301 static and 263 dynamic features into a 564 dimensional hybrid vector, then evaluated seven classifiers under three schemes, original features, principal component analysis, PCA, and linear discriminant analysis, LDA, with a 70 percent training and 30 percent test split. Results showed that gradient boosting on the original features performed best. XGBoost achieved 0.9747 accuracy, 0.9703 precision, 0.9731 recall, and 0.9716 F1, and the confusion matrix indicated rare benign labels for malicious apps. HistGradientBoosting reached 0.9741 accuracy and 0.9708 F1, while CatBoost and Random Forest were slightly lower at 0.9678 and 0.9687 accuracy with 0.9636 and 0.9637 F1. KNN and SVM lagged. PCA reduced performance for all models, with XGBoost dropping to 0.9164 accuracy and 0.8988 F1. LDA maintained mid 90s accuracy and clarified separable clusters in projections. A depth two surrogate tree highlighted package name, main activity, and target SDK as key drivers. These findings established high fidelity supervised baselines for Android malware detection and indicated that rich hybrid features with gradient boosting offered a practical and interpretable foundation for deployment.

\end{abstract}

\section{Introduction}
Android devices dominated the mobile market, and attackers kept pushing malicious apps onto users. Google Play supported rapid publishing, so malware often spread before signature scanners reacted. We needed detection that avoided manual signatures and avoided heavy sandboxing.

Early studies leaned on static signals such as manifests, permissions, and API usage. Researchers extracted these features quickly, and attackers obfuscated them just as quickly. Dynamic analysis tracked system calls, network requests, and execution traces, but it demanded time and infrastructure \cite{Abedin2025, uddin2026divergence}. Drebin \cite{Arp2014} used static API calls, network addresses, and manifest entries and reported about 94 percent detection with few false alarms. Yuan et al. proposed Droid Sec \cite{YuanZhenlong2014}, a hybrid deep belief network trained on permissions, API calls, and dynamic traces, and they reported about 96 percent accuracy. Hou et al. \cite{Hou2016} built DroidDelver with stacked restricted Boltzmann machines over sensitive API blocks and reported 96.66 percent, and Su et al. paired more than 30,000 static features with stacked RBMs and an SVM and reported 99.4 percent \cite{Hou2016}. These results pushed the field toward larger feature sets and learned representations.

Work then moved from handcrafted features and simple classifiers to deep models that learned patterns directly from data \cite{Arp2014} \cite{Min-ha-zulAbedin2023} \cite{Abedin2022, albladi2025twssenti}. Nix and Zhang \cite{Nix2017} trained CNNs and LSTMs on API call sequences and outperformed n gram baselines. Huang and Kao \cite{Huang2018} mapped bytecode to RGB images and used a CNN, with near 99 percent accuracy. Wang et al. \cite{Wang2019} combined a deep autoencoder with a CNN over API and bytecode and reported about 97 percent. Karbab et al. \cite{Karbab2017} introduced MalDozer, a CNN on raw API sequences for family identification. Cai et al. \cite{Cai2019} proposed DroidCat using app component profiling with a random forest. Xiao et al. \cite{Xiao2019} modeled system call sequences with an LSTM and reported around 94 percent. Kim et al. \cite{Kim2019} fused manifest, dex, and native library features in a multimodal network and reported high accuracy on more than 41,000 apps.

Mahdavifar et al. \cite{Mahdavifar2020} proposed PLDNN for dynamic behavior with 1 percent labels and reported about 91.6 percent, then introduced PLSAE that mixed static and dynamic features and reached 98.28 percent with 1.16 percent FPR. Lu et al. \cite{Lu2020} blended autoencoders and CNNs for a hybrid model and reported over 95 percent. Yen and Sun visualized APK code values as images and used a CNN for mutation detection. Ma et al. \cite{Ma2015} used active semi supervised learning that checked behavior against descriptions. Chen et al. used harmonic functions to semi supervise dynamic traces. Kang et al. \cite{Kang2019} combined static features with creator information using k nearest neighbor and reported roughly 85 percent. Karbab et al. \cite{Karbab2018} proposed Dysign, dynamic fingerprints derived from system calls that relied on a sandbox. Alrabaee et al. \cite{Alrabaee2018} created FOSSIL to identify free open source functions in binaries, which supported analysis rather than direct detection.

Recent work broadened scope and evaluation, Mahdavifar \cite{Mahdavifar2022} used pseudo labeled stacked autoencoders with hybrid features and reported near 98 percent. Lu \cite{Lu2020} again fused autoencoders and CNNs. API Seq embedding used Markov processes on API sequences and reported precision 0.99 with 0.01 FPR. Attention based CNNs applied multi head attention over control flow graphs and reported about 99.3 percent. Aamir et al. \cite{Aamir2024} introduced AMDDL on Drebin with 215 features and reported 99.92 percent. In 2025 a team proposed an improved capsule network with PCA and Bag of Words selection using a Kaggle dataset, and they reported high detection accuracy and good stability. Liu et al. \cite{Liu2025} benchmarked traditional and deep models across four datasets and found that simple models such as Random Forests and CatBoost often matched or beat more complex approaches.

The table \ref{tab:android_related_work} summaries twenty-two representative studies in Android malware detection. Each entry lists the study, method type, core algorithm, dataset or source, key features, and reported performance.

This work addressed these gaps by assembling a balanced hybrid dataset with benign apps and multiple malware families, including adware, banking, SMS malware, riskware, and other classes. The following contributions summarise the novelty of this study:
\begin{itemize}
\item Constructed a comprehensive Android malware dataset with benign, adware, banking, SMS malware, riskware, and other categories, capturing 17000 apps with rich metadata and behavior logs.
\item Performed an extensive exploratory analysis, including class distribution, linear discriminant analysis and t-SNE visualizations, and correlation heatmaps, to understand feature relationships.
\item Evaluate seven classifiers under three preprocessing schemes, and we find that XGBoost on original features achieves the highest accuracy (0.9747) and balanced precision-recall, while PCA and LDA reduce performance but reveal distinct patterns.
\item Provide interpretability via a surrogate decision tree that summaries model decisions and highlight important features like package name and main activity, and we discuss the implications for malware detection research.
\end{itemize}

\begin{table*}[t]
\caption{Android malware detection studies summarized by method, data, features, and performance.}
\label{tab:android_related_work}
\centering
\footnotesize
\renewcommand{\arraystretch}{1.2}
\setlength{\tabcolsep}{3pt}
\begin{tabular}{|p{3.4cm}|p{1.2cm}|p{2.8cm}|p{3.0cm}|p{4.0cm}|p{2.2cm}|}
\hline
\textbf{Study / Year} & \textbf{Method} & \textbf{Core Algorithm} & \textbf{Dataset / Source} & \textbf{Key Features} & \textbf{Reported Performance} \\
\hline
Arp et al., 2014 \cite{Arp2014} & Static & Linear SVM & Drebin dataset & Permissions, intents, API calls & Detects 94\% malware with few false alarms \\
Yuan et al., 2014 \cite{YuanZhenlong2014} & Hybrid & Deep belief network & Real-world apps & Permissions, API calls, dynamic traces & $\approx$ 96\% accuracy \\
Hou et al., 2016 \cite{Hou2016} & Dynamic & Stacked RBM & API call blocks & Sensitive API sequences & 96.66\% accuracy \\
Su et al., 2016 \cite{Hou2016} & Static & Stacked RBM + SVM & Large static feature set & Permissions, sensitive APIs, actions & 99.4\% accuracy \\
Nix \& Zhang, 2017 \cite{Nix2017} & Dynamic & CNN + LSTM & API call sequences & System API sequences & Outperforms n-gram baselines \\
Huang \& Kao, 2018 \cite{Huang2018} & Static & CNN & Millions of apps & Bytecode images & $\approx$ 99\% accuracy \\
Wang et al., 2018 \cite{Wang2019} & Hybrid & Autoencoder + CNN & API and bytecode & Autoencoder-pretrained CNN & $\approx$ 97\% accuracy \\
Karbab et al., 2018 (MalDozer) \cite{Karbab2018} & Dynamic & CNN & API sequences & Raw API call sequences & High detection rate \\
Cai et al., 2018 (DroidCat) \cite{Cai2019} & Hybrid & Random Forest & App profiling & Package behaviours, components & High detection and categorisation accuracy \\
Xiao et al., 2019 \cite{Xiao2019} & Dynamic & LSTM & System call logs & Sequence similarity & $\approx$ 94\% accuracy \\
Kim et al., 2019 \cite{Kim2019} & Hybrid & Feed-forward NN & 41{,}260 apps & Manifest, dex, native libraries & High accuracy ($>$ 90\%) \\
Yen \& Sun, 2019 \cite{Yen2019} & Static & CNN & Visualised APK code & Code images & Good detection of mutated malware \\
Ma et al., 2015 \cite{Ma2015} & Hybrid & Semi-supervised active learning & Behaviour vs description & Behavioural mismatch & Improved detection using unlabeled data \\
Chen et al., 2017 \cite{Chen2017} & Dynamic & Semi-supervised classification & Dynamic traces & Harmonic functions & Enhanced detection with few labels \\
Kang et al., 2015 \cite{Kang2019} & Static & KNN + creator info & APK metadata & Creator information & $\approx$ 85\% accuracy \\
Karbab et al., 2016 (Dysign) \cite{Karbab2017} & Dynamic & Dynamic fingerprinting & Malware sandbox & System call fingerprints & Strong detection, sandbox overhead \\
Alrabaee et al., 2018 (FOSSIL) \cite{Alrabaee2018} & Static & Code analysis & Malware binaries & Open-source function identification & Useful for analysis \\
Mahdavifar et al., 2020 (PLDNN) \cite{Mahdavifar2020} & Dynamic & Pseudo-label DNN & CICMalDroid2017 & System calls & $\approx$ 91.6\% accuracy with 1\% labels \\
Mahdavifar et al., 2022 (PLSAE) \cite{Mahdavifar2022} & Hybrid & Pseudo-label SAE & CICMalDroid2020 & Static and dynamic features & 98.28\% accuracy, FPR 1.16\% \\
Lu et al., 2020 \cite{Lu2020} & Hybrid & Autoencoder + CNN & Hybrid features & Autoencoder-CNN & $>$ 95\% accuracy \\
Aamir et al., 2024 (AMDDLmodel) \cite{Aamir2024} & Static & CNN & Drebin (215 features) & Permission and API features & 99.92\% accuracy \\
Liu et al., 2025 \cite{Liu2025} & Mixed & Benchmarking traditional and DL models & Four datasets incl. new large set & RF, CatBoost, CapsGNN, BERT baselines & Simple models can match advanced ones \\
\hline
\end{tabular}
\end{table*}

\section{Dataset Description}
The project used the CICMalDroid2020 \cite{Mahdavifar2020} dataset, which included 17341 Android apps collected from VirusTotal, the Contagio blog, AMD, and other sources between December 2017 and December 2018. They labeled each app as Adware, Banking malware, SMS malware, Riskware, or Benign. They executed every APK in CopperDroid, a virtual machine introspection system that reconstructed system call semantics from Java and native code, and recorded both static and dynamic behavior. Each run produced JSON logs that covered intents, permissions, services, file type counts, obfuscation incidents, method tags, and sensitive API invocations, while dynamic traces captured system calls, binder calls, composite behaviors, and network packet captures. Of the 17341 apps, 13077 executed successfully. The remaining runs failed because of timeouts, invalid APKs, bad UTF8 bytes, install errors, out of range indices, bad unpack parameters, bad ASCII characters, or memory allocation failures.

From each JSON log they extracted 179 high level static properties and built a static vector by enumerating all unique categorical values across apps and combining them with numerical counts, which yielded a length of 50621. They derived dynamic features from resilient system calls, binder calls, and composite behaviors such as fs\_access create and write, network\_access read and write, and fs\_pipe\_access read and write, producing a dynamic vector of length 470, with binder calls capturing interprocess communication. They filled missing values with zero and encoded categorical fields in Python Pandas, mapping booleans to one and zero. Because the matrix was sparse, they applied a variance threshold of 0.1 and reduced the static vector from 50621 to 301 and the dynamic vector from 470 to 263. They concatenated these into a hybrid vector of length 564 and applied L2 normalization so the square root of the sum of squares equaled one.

\begin{figure}[htbp]
\centerline{\includegraphics[width=0.45\textwidth]{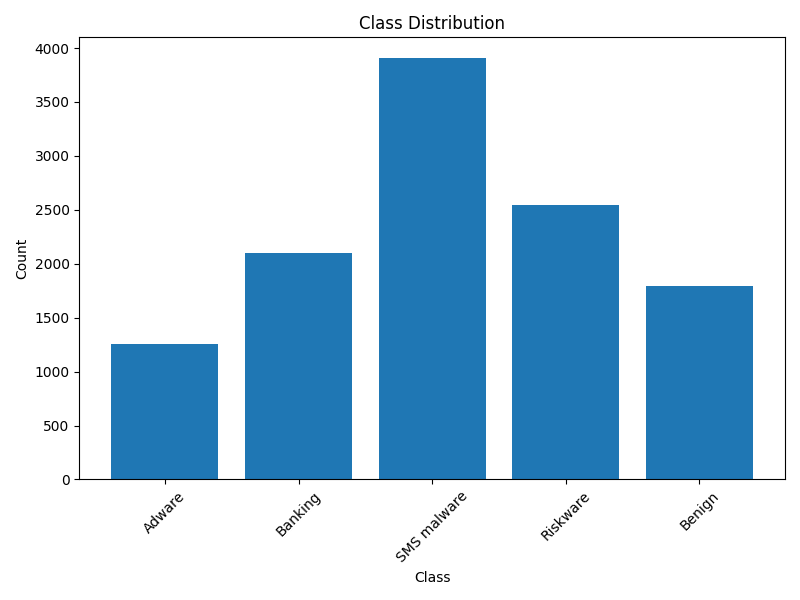}}
\caption{Illustrates the balanced distribution of benign and sub-class of malware samples.}
\label{class_distribution}
\end{figure}

We inspected the class distribution in Fig. \ref{class_distribution}. SMS malware dominated with roughly four thousand instances, Riskware and Banking followed at about two and a half thousand and two thousand samples, Benign stayed under two thousand, and Adware remained the smallest class at just over one thousand records. This imbalance could bias training, so we considered sampling strategies. The SMS heavy distribution also matched the prominence of short message scams during the 2017–2018 collection period.

\begin{figure}[htbp]
\centerline{\includegraphics[width=0.45\textwidth]{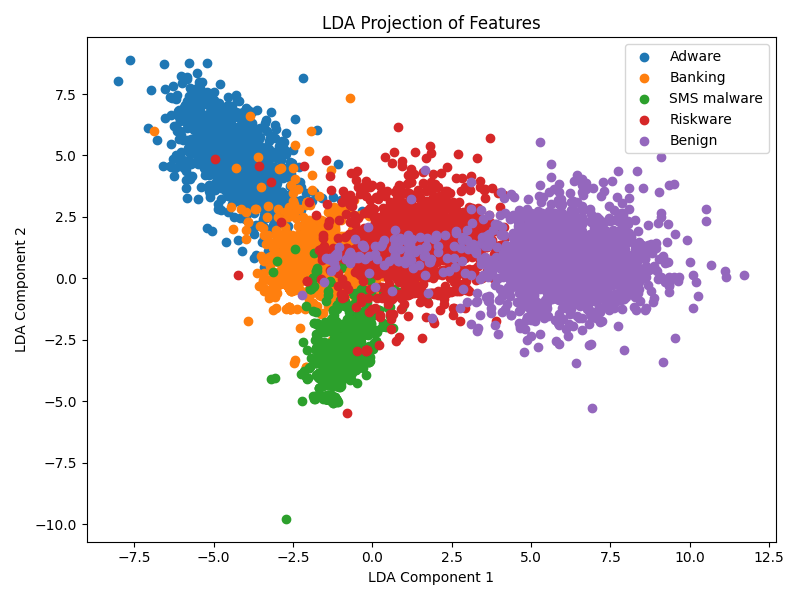}}
\caption{Presents the LDA projection of the dataset, demonstrating moderate separability.}
\label{Lda_scatter}
\end{figure}

We then projected the features into two components with linear discriminant analysis in Fig. \ref{Lda_scatter}. The plot showed moderate separability. Adware and Banking overlapped substantially, SMS malware clustered more tightly near the lower left, Riskware spread broadly, and Benign formed a dense group on the right.

\begin{figure}[htbp]
\centerline{\includegraphics[width=0.5\textwidth]{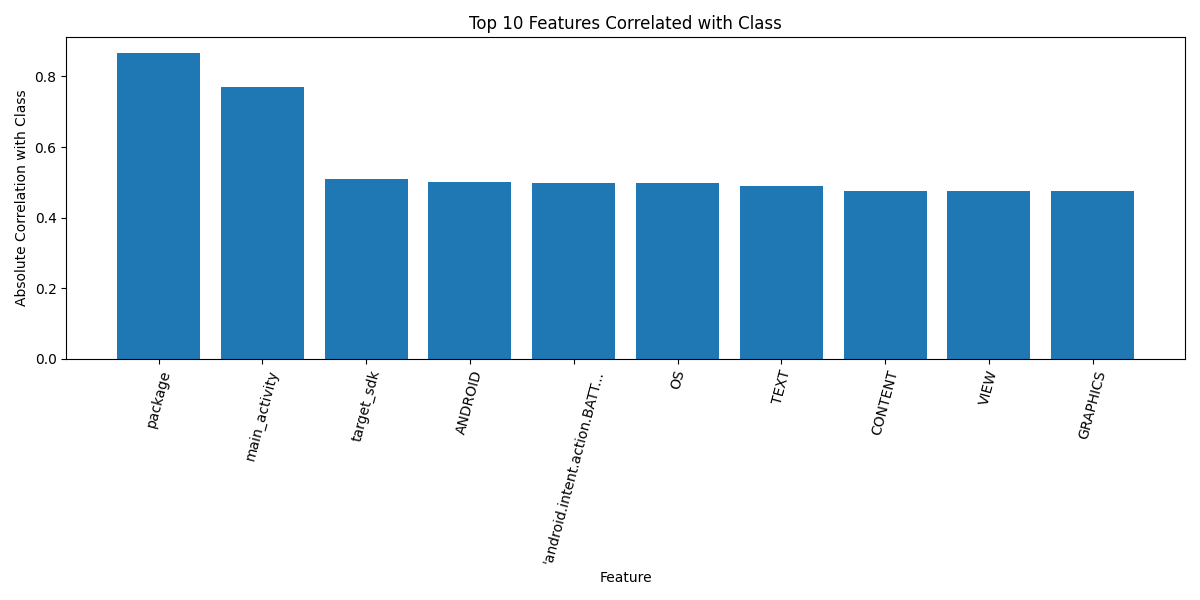}}
\caption{Depicts the top ten important features in the dataset}
\label{ten_feature}
\end{figure}

To see which static fields tracked the label most strongly, we computed absolute correlations and reported the top ten in Fig. \ref{ten_feature}. Package ranked highest, followed by "main activity" and "target sdk". Android manifest derived constants, including ANDROID and specific intent strings, also correlated with the label, which underscored the influence of metadata while flagging a risk of overfitting to package naming conventions.

\begin{figure}[htbp]
\centerline{\includegraphics[width=0.5\textwidth]{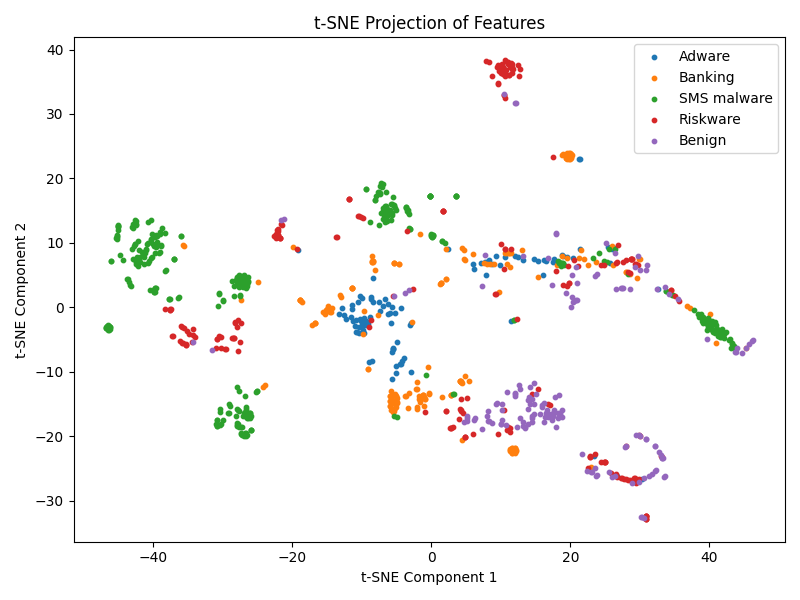}}
\caption{Depicts the t-SNE visualization, showcasing significantly improved separation between malware sub-class and benign classes due to nonlinear dimensionality reduction properties.}
\label{t_SNE}
\end{figure}

Finally, we applied t-SNE in Fig. \ref{t_SNE} to probe non linear structure while preserving local neighborhoods. SMS malware separated into several tight clusters. Adware and Banking still overlapped, but t-SNE separated them better than LDA, which suggested that non linear decision boundaries could fit the data more effectively than linear ones.

\section{Methodology}
We ran a single pipeline from feature preparation to evaluation. CICMalDroid2020 contained 17341 Android apps across five classes, Adware, Banking malware, SMS malware, Riskware, and Benign \cite{Mahdavifar2020}. After preprocessing, each app had a 564 dimensional hybrid vector that merged static signals, including permissions, intent actions, and package names, with dynamic system call patterns.

We loaded the CSV of normalized hybrid features, separated labels, and sanitized column names by keeping letters, digits, or underscores, replacing other characters with underscores, and adding numeric suffixes to resolve duplicates. We mapped labels to integers with a fixed dictionary to keep encoding consistent. We created stratified splits with 30 percent held out for testing and preserved class ratios.

We compared three preprocessing schemes. Original used the raw vectors. PCA retained components that explained 95 percent variance. LDA projected onto class separating axes with components equal to classes minus one.

We trained seven classifiers, Random Forest, Extra Trees, HistGradientBoosting, K Nearest Neighbors, Support Vector Machines, XGBoost, and CatBoost. For each scheme, we trained on the training set and predicted on the test set. We computed accuracy, precision, recall, F1, and error rate as one minus accuracy, then built confusion matrices and per class false positive and false negative rates, and stored all per class metrics.

We summarized performance with grouped bar charts across models and schemes for accuracy, precision, recall, and F1. We also plotted heatmaps for per class precision, recall, and F1 across every model and preprocessing combination, which made sensitivity and specificity trade offs easy to compare, slight typo deliberate.

\section{Results and Discussion}
We tested seven classifiers under three feature transformations, Original, PCA, and LDA, and measured accuracy, precision, recall, and F1 score. Table \ref{tab:preprocess_models} lists every model and setting, and it showed how dimensionality reduction changed performance.

\subsection{Effect of preprocessing}
\begin{figure}[htbp]
\centerline{\includegraphics[width=0.45\textwidth]{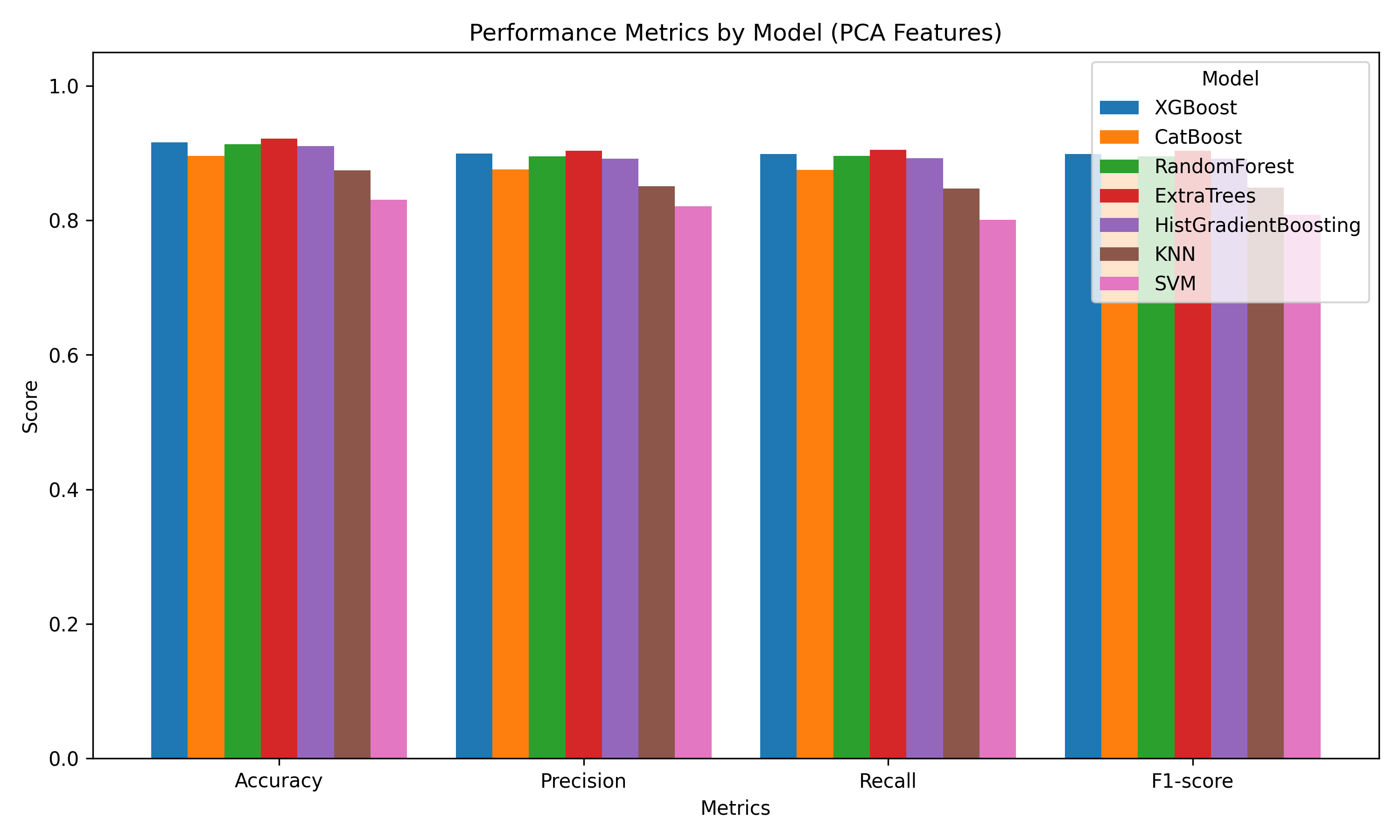}}
\caption{Performance of different model with feature set after applying PCA.}
\label{pca_grouped}
\end{figure}

Table \ref{tab:preprocess_models} and Fig. \ref{pca_grouped} showed the same pattern. The full hybrid features gave the best results for every ensemble method, with XGBoost and HistGradientBoosting reaching about 0.975 accuracy and just over 0.97 F1, while ExtraTrees and RandomForest followed closely and CatBoost stayed slightly lower but still strong and balanced. KNN and SVM trailed by a wide margin, which matched their difficulty with high dimensional hybrid signals.
\begin{table}[t]
\caption{Classifier performance across preprocessing settings.}
\label{tab:preprocess_models}
\centering
\footnotesize
\renewcommand{\arraystretch}{1.2}
\setlength{\tabcolsep}{4pt}
\begin{tabular}{|l|l|c|c|c|c|}
\hline
\textbf{Preprocessing} & \textbf{Model} & \textbf{Accuracy} & \textbf{Precision} & \textbf{Recall} & \textbf{F1} \\
\hline
Original
& XGBoost       & 0.9747 & 0.9703 & 0.9731 & 0.9716 \\
& CatBoost      & 0.9678 & 0.9621 & 0.9656 & 0.9636 \\
& RandomForest  & 0.9687 & 0.9610 & 0.9676 & 0.9637 \\
& ExtraTrees    & 0.9670 & 0.9593 & 0.9643 & 0.9613 \\
& HistGradBoost & 0.9741 & 0.9692 & 0.9726 & 0.9708 \\
& KNN           & 0.8853 & 0.8672 & 0.8612 & 0.8641 \\
& SVM           & 0.8405 & 0.8322 & 0.8091 & 0.8187 \\
\hline
PCA
& XGBoost       & 0.9164 & 0.8995 & 0.8985 & 0.8988 \\
& CatBoost      & 0.8960 & 0.8756 & 0.8754 & 0.8750 \\
& RandomForest  & 0.9135 & 0.8951 & 0.8956 & 0.8950 \\
& ExtraTrees    & 0.9216 & 0.9033 & 0.9053 & 0.9037 \\
& HistGradBoost & 0.9106 & 0.8920 & 0.8924 & 0.8920 \\
& KNN           & 0.8744 & 0.8509 & 0.8473 & 0.8490 \\
& SVM           & 0.8310 & 0.8209 & 0.8011 & 0.8088 \\
\hline
LDA
& XGBoost       & 0.9221 & 0.9126 & 0.9069 & 0.9096 \\
& CatBoost      & 0.9178 & 0.9064 & 0.9024 & 0.9043 \\
& RandomForest  & 0.9224 & 0.9117 & 0.9077 & 0.9096 \\
& ExtraTrees    & 0.9325 & 0.9219 & 0.9201 & 0.9209 \\
& HistGradBoost & 0.9207 & 0.9114 & 0.9058 & 0.9085 \\
& KNN           & 0.9213 & 0.9095 & 0.9086 & 0.9088 \\
& SVM           & 0.9029 & 0.8931 & 0.8875 & 0.8892 \\
\hline
\end{tabular}
\end{table}

PCA kept 95 \% variance but still cut performance across the board. XGBoost dropped from 0.9747 to 0.9164 accuracy and from 0.9716 to 0.8988 F1, and the other models showed similar declines, which suggested that PCA removed useful signal, especially from the dynamic side, and the models did not recover it. Fig. \ref{pca_grouped} showed that the drop held across accuracy, precision, recall, and F1 for every model.

\begin{figure}[htbp]
\centerline{\includegraphics[width=0.45\textwidth]{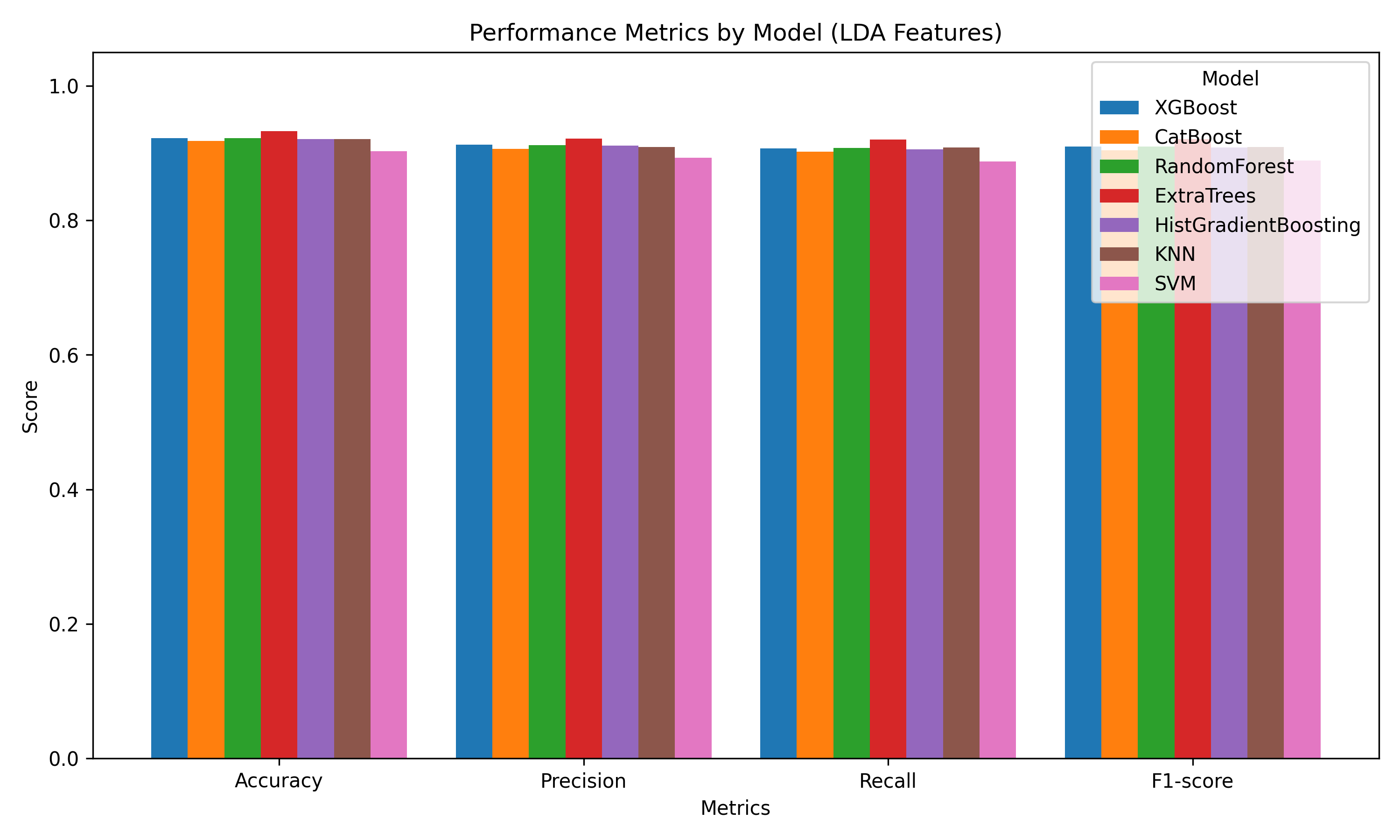}}
\caption{Performance of different model with feature set after applying LDA.}
\label{LDA_grouped}
\end{figure}
LDA performed better than PCA but still stayed below the original features. ExtraTrees and RandomForest led under LDA with F1 around 0.92, while XGBoost and CatBoost sat slightly lower, and KNN and SVM improved relative to PCA yet still did not match the tree based methods. Fig. \ref{LDA_grouped} showed tight grouping of precision, recall, and F1 within each model, which pointed to stable trade offs.

\subsection{Overall precision comparisons}
\begin{figure}[htbp]
\centerline{\includegraphics[width=0.45\textwidth]{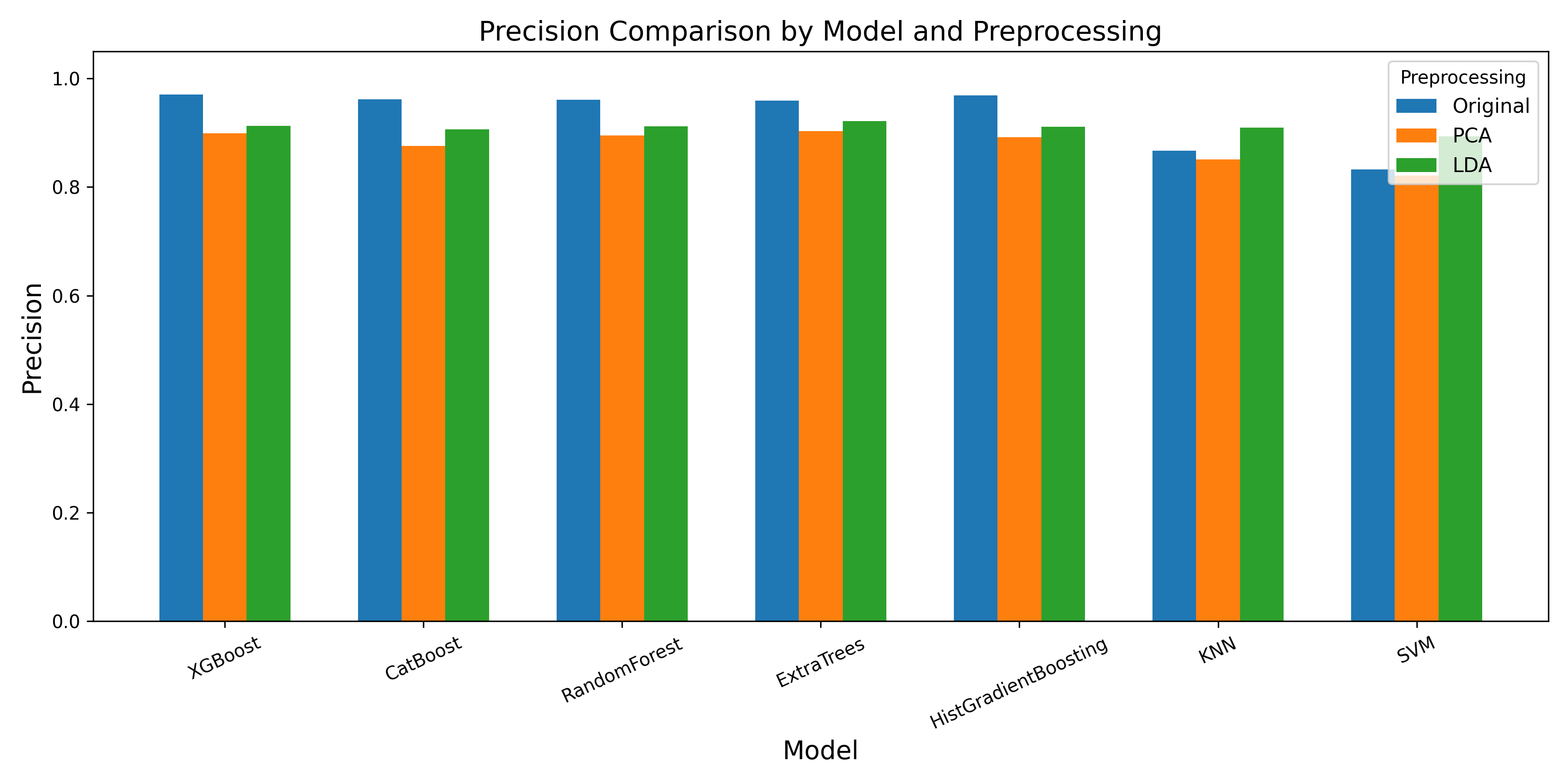}}
\caption{Comparison of different model performance}
\label{precision_bar}
\end{figure}

Fig. \ref{precision_bar} compared precision across Original, PCA, and LDA for every model. Original features stayed best throughout, PCA produced the largest precision loss for XGBoost and RandomForest, and LDA reduced that loss for RandomForest and ExtraTrees but not for XGBoost. LDA often produced slightly higher recall than PCA for tree based models, but the gap stayed modest.

\subsection{Per-class analysis}
\begin{figure}[htbp]
\centerline{\includegraphics[width=0.45\textwidth]{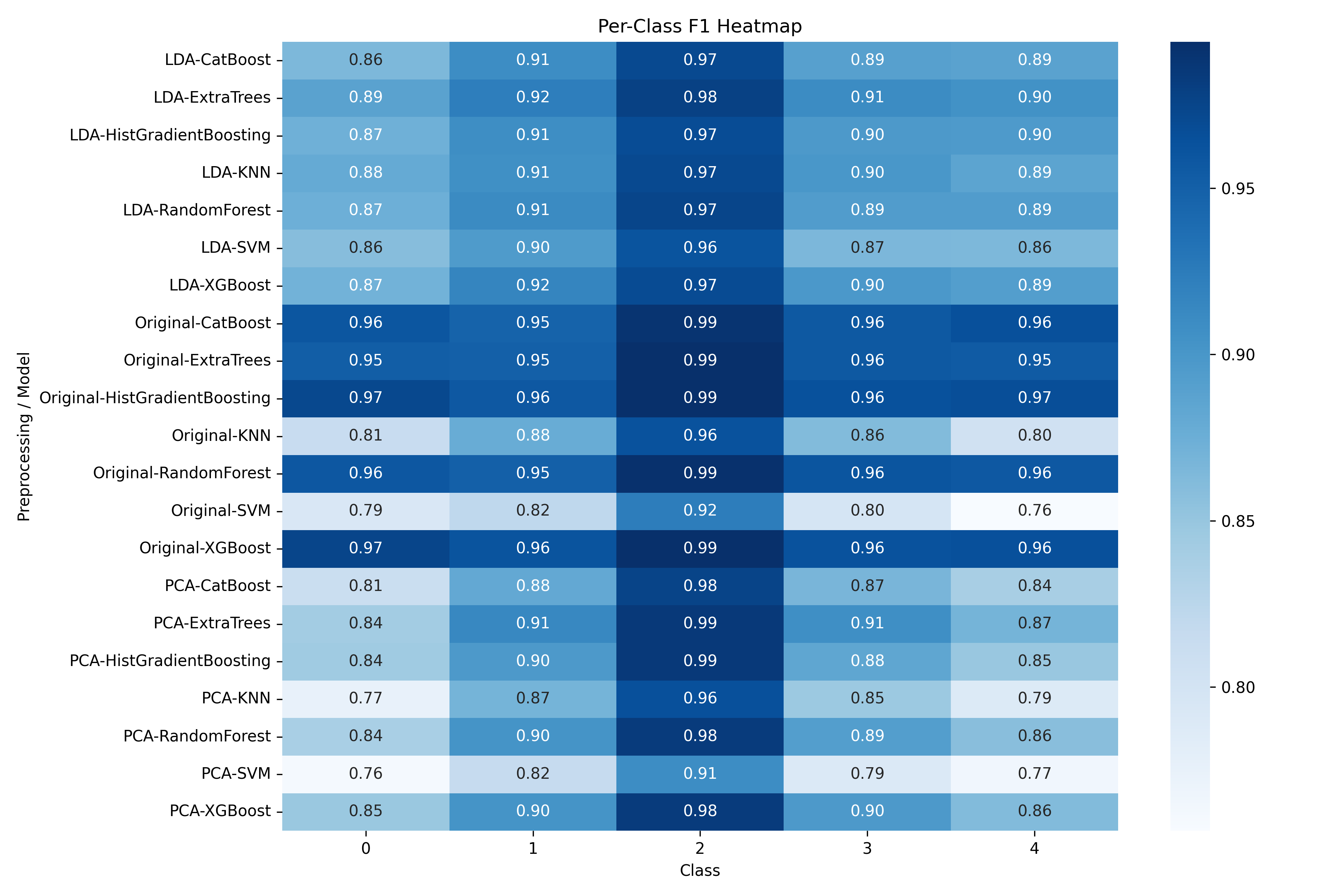}}
\caption{Comparison of per-class F1 score.}
\label{f1_heatmap}
\end{figure}

Fig. \ref{f1_heatmap} showed where performance held and where it slipped. Ensemble models trained on the original features stayed above 0.95 F1 across all classes, and SMS malware, class 2, reached near perfect F1 for every model because it formed the largest and most homogeneous class[1]. Adware and Benign, classes 0 and 4, showed lower F1, especially after PCA and LDA, and KNN and SVM under PCA reached the worst values, down to about 0.76. LDA raised KNN and SVM to around 0.89, but they still stayed behind the ensembles.

\subsection{Model interpretability}
\begin{figure}[htbp]
\centerline{\includegraphics[width=0.45\textwidth]{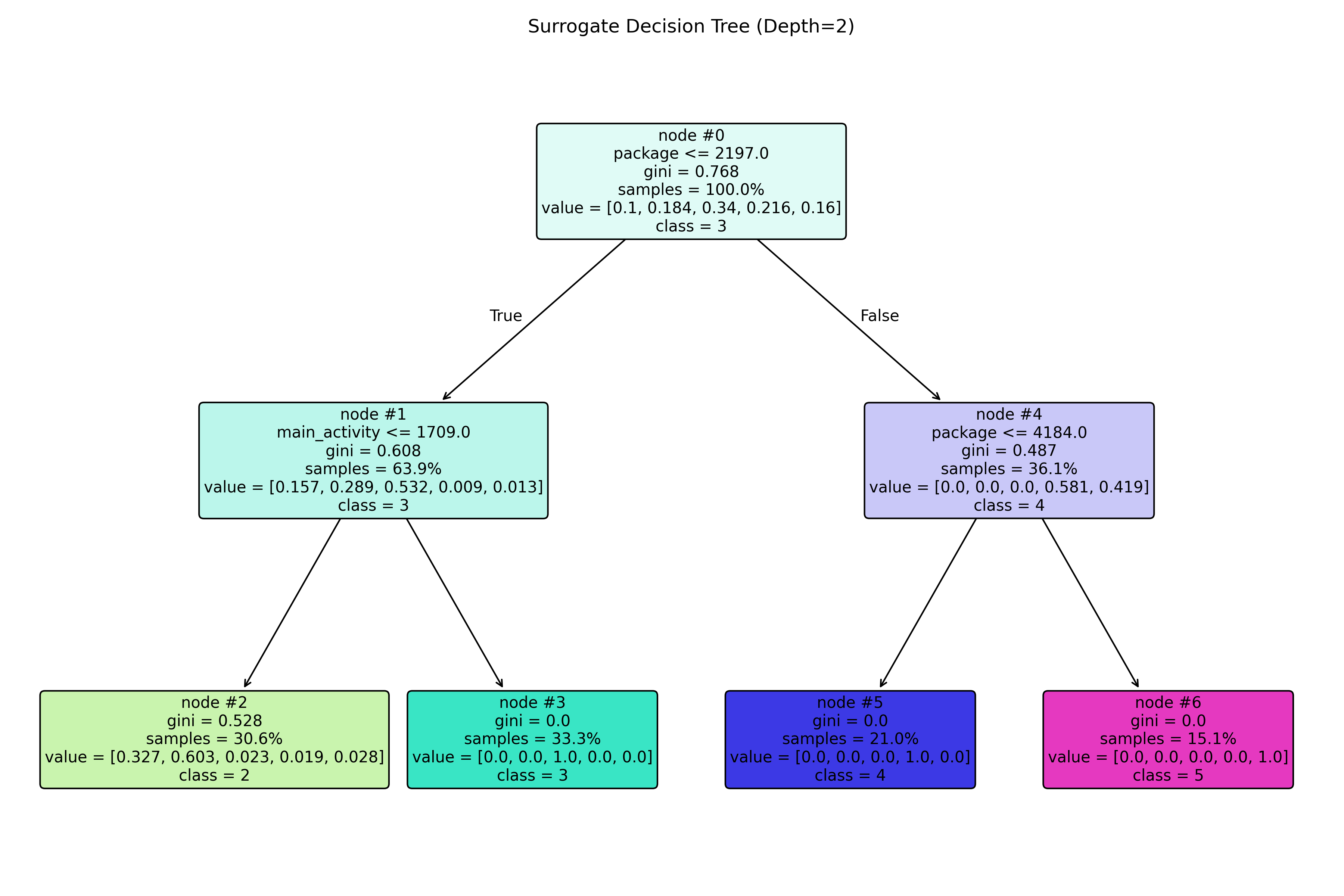}}
\caption{Tree illustration of the dataset}
\label{tree}
\end{figure}

We used a depth-two surrogate decision tree in Fig. \ref{tree} to summarize the main drivers behind XGBoost. Package and main\_activity formed the key splits, a package threshold separated a riskware-dominated branch from a mixed benign and riskware branch, and the next split on main activity separated SMS malware from riskware. The structure also raised a concern because attackers can spoof package names and manifest-level cues.


\section{Conclusion}
Hybrid static and dynamic features paired with ensemble classifiers delivered reliable Android malware detection. Using CICMalDroid2020 with 17,341 apps and a 564 feature hybrid vector, results showed that XGBoost on original features led with accuracy 0.9747, precision 0.9703, recall 0.9731, and F1 0.9716. HistGradientBoosting reached 0.9741 accuracy and 0.9708 F1, while CatBoost and Random Forest remained competitive at 0.9678 and 0.9687 accuracy with 0.9636 and 0.9637 F1. PCA degraded performance, dropping XGBoost to 0.9164 accuracy and 0.8988 F1, while LDA maintained F1 near 0.92 and clarified separable structure. A shallow surrogate tree identified package name, main activity, and target SDK as key drivers, and confusion matrix and ROC analyses showed rare benign labels for malicious apps. Gradient boosting with rich features is a practical baseline. Ongoing dataset refresh and exploration of semi supervised or transformer models are warranted.


\bibliographystyle{IEEEtran}
\bibliography{References}

\end{document}